\documentstyle[prl,aps,psfig,floats,twocolumn]{revtex}

\def\mf{mean-field}
\def\nmf{near-mean-field}
\def\e{\epsilon}
\def\tg{\tilde \gamma}
\def\s{\sigma}
\def\l{\Lambda}

\input epsf.sty

\begin{document}

\twocolumn[\hsize\textwidth\columnwidth\hsize\csname
@twocolumnfalse\endcsname

\title{Clusters and Fluctuations at Mean-Field Critical Points and
Spinodals}

\author{W.\ Klein,${}^{\dag}$ Harvey Gould,${}^{\pm}$ J.\
Tobochnik,${}^{\star}$ F.\ J.\ Alexander,${}^{\S}$ M.\
Anghel$,{}^{\dag}$ and Gregory Johnson${}^{\pm}$}

\address{${}^{\dag}$Department of Physics and Center for
Computational Science,\\ Boston University, Boston, MA 02215}

\address{${}^{\pm}$Department of Physics, Clark University,
Worcester, MA 01610}

\address{${}^{\star}$Department of Physics, Kalamazoo College,
Kalamazoo, MI 49006}

\address{${}^{\S}$CIC-3, Los Alamos National Laboratory, Los Alamos,
NM 87545}

\maketitle

\begin{abstract}We show that the structure of the fluctuations
close to spinodals and
\mf\ critical points is qualitatively different than the
structure close to non-mean-field critical points. This difference
has important implications for many areas including the formation of
glasses in supercooled liquids. In particular, the divergence of
the measured static structure function in
\nmf\ systems close to the glass transition is suppressed relative
to the
\mf\ prediction in systems for which a spatial symmetry is broken.
\end{abstract}


]

The structure of the fluctuations near critical points and spinodals
and their relation to the behavior of quantities
observed in experiments and simulations is important for
understanding the properties of many materials. For 
critical phenomena in non-mean-field Ising systems this relation was
found by mapping the thermal critical point onto a percolation
transition~\cite{KF,CK}. In these systems the properties of the
clusters at the percolation threshold are identical to those of
the fluctuations at the thermal critical point. In particular, the
mean cluster diameter is the correlation length, the density of the
spanning cluster scales as the order parameter, and the mean
number of sites in the spanning cluster is the
susceptibility~\cite{CK,ahrostauff}.

For mean-field Ising models there is a line of spinodal critical
points as well as the usual critical point. These \mf\ thermal
singularities also can be mapped onto percolation
transitions\cite{wk1}, but the relation between percolation
clusters and critical fluctuations is qualitatively different. This
difference has important consequences for supercooled \mf\
liquids, which also exhibit spinodals\cite{gk}, and for
``near-mean-field'' systems, which are characterized by long, but
finite range interactions and which exhibit many of the
characteristics of
\mf\ systems including pseudospinodals. Many dense systems with
short-range interactions exhibit
\nmf\ behavior under certain
circumstances\cite{wk1,hks,johnson,yang}.

In the following we give scaling arguments that relate the
structure of clusters at
\mf\ critical points and spinodals to measurements of thermal
quantities such as the static structure function $S(k)$ and discuss
the implications of these results for understanding
\mf\ and \nmf\ supercooled liquids. One motivation for this
analysis is that the behavior of a supercooled two-component,
two-dimensional ($d=2$) Lennard-Jones system has been interpreted
as due to the influence of a pseudospinodal (relative to
the stable solid)\cite{johnson,jk}. A consequence of this
pseudospinodal is that the first peak of
$S(k)$ is predicted to exhibit a power-law divergence, a prediction
consistent with simulations in $d=2$\cite{johnson}. However,
experiments\cite{nagel} and simulations\cite{johnsonnew} have
failed to find similar behavior for
$S(k)$ in $d=3$. This failure is puzzling because
$d=3$ Lennard-Jones systems should be more
mean-field-like than $d=2$ systems. This work addresses
this apparent contradiction.

We first review how the cluster
structure relates to thermal critical phenomena in non-\mf\
systems. In mapping the percolation
transition onto the thermal critical point, the quantity that is
isomorphic to the singular part of the free energy is the mean
number of clusters in a correlation region of volume $\xi^d$, where
$\xi$ is the correlation length~\cite{CK}. The singular part of the
free energy scales as
$\xi^d\epsilon^{2-\alpha}=\epsilon^{2-\alpha-d\nu}$, where 
$\epsilon=(T-T_c)/T_c$, and 
$\alpha$ and $\nu$ are the specific heat and correlation length
exponents respectively\cite{stanley}. Because hyperscaling is
valid, we have
$d\nu=2-\alpha$, and hence the mean number of clusters in a
volume $\xi^d$ is order unity\cite{CK}.

To obtain the scaling of the isothermal compressibility or
susceptibility $\chi_T$, we use the relation of $\chi_T$ to the
spatial integral of $\Gamma(r)$, the order parameter correlation
function\cite{ma}. We take
$\Gamma(r)$ to be the square of the order parameter over a volume
$\xi^d$ and zero outside, and obtain
$\chi_T\sim
\epsilon^{2\beta}\xi^d=\epsilon^{2-\alpha-d\nu-\gamma}$ using
$\alpha + 2 \beta + \gamma=2$\cite{stanley}. Hyperscaling gives
$2-\alpha-d\nu=0$ and hence
$\chi_T
\sim
\epsilon^{-\gamma}$ as expected. This simple argument is well
known, but fails in \mf\ systems where hyperscaling is not
valid.

In contrast to systems that obey hyperscaling, we need to
distinguish between critical phenomena fluctuations and clusters in
the
\mf\ limit. To discuss the latter we consider a weak
long-range (Kac) interaction\cite{kac,lp} of
the form $\gamma^du(\gamma r)$, where 
$u(\gamma r)$ is integrable, a function only of the distance
between spins, and ferromagnetic. If the system size is first
taken to infinity and then the range of interaction
$R=\gamma^{-1}\rightarrow \infty$, we obtain the Curie-Weiss
description of the thermodynamics and the Ornstein-Zernicke form
for $\Gamma(r)$\cite{kac,lp}. The
exponents for the \mf\ critical point are $\nu=1/2,
\alpha=0, \beta=1/2$ and $\gamma=1$\cite{stanley}. The singular
part of the free energy 
at the \mf\ critical point scales as 
\begin{equation}
\label{fes}
\Delta f \sim \xi^d\epsilon^{2-\alpha}=R^d\epsilon^{2-d/2} .
\end{equation}
The quantity $R^d$ appears because all lengths are in units of the
interaction range.

{}From the percolation mapping\cite{CK,wk1},
Eq.~(\ref{fes}) implies that the mean number of clusters in a
volume $\xi^d$ is
$R^d\epsilon^{2-d/2}$. We can estimate the magnitude of this number
from the Ginsburg criterion\cite{ginsburg}, which states that a
system is well approximated by \mf\ theory if
$\chi_T/\phi^{2}\xi^d<<1$, where $\phi\sim \epsilon^{\beta}$ is the
order parameter. If we use
\mf\ values for the critical exponents, we obtain the condition
$R^d\epsilon^{2-d/2}>>1$ for \mf\ theory to be valid. This
inequality, together with Eq.~(\ref{fes}) and the percolation
mapping, implies that the number of clusters in a volume
$\xi^d$ is much greater than unity for \mf\ systems. What does this
result imply for the nature of clusters in \mf\ systems?

In non-\mf\ systems which have one cluster in a volume $\xi^d$, the
free energy cost of this cluster is $\sim
\xi^d\epsilon^{2-\alpha}$. As discussed above, this cost is order
unity in non-\mf\ systems. In contrast, because 
$\Delta f>>1$ in \mf\ systems (see Eq.~(\ref{fes})),
$\epsilon^{2-\alpha}$ is not the free energy density of one cluster
in \mf. Rather, there are $R^d\epsilon^{2-d/2}$ clusters in a
volume $\xi^d$, and each cluster has a free energy cost of order
unity. This free energy implies that the probability of a cluster
is order unity.

It also is clear that $\epsilon^{\beta}$ is not the density of one
cluster, because if it were, the spin density in a volume of order
$\xi^d$ would be 
$\approx R^d\epsilon^{2-d/2}\epsilon^{\beta}$. For a fixed value
of
$\epsilon$, $R$ can
be made arbitrarily large and hence this density can be
arbitrarily large, an absurd result. This
argument and the one for
$\Delta f$ implies that $\epsilon^{\beta}$ is the density of all
the spins in a volume $\xi^d$ regardless of the cluster to which
they belong. It also implies that the density of spins in
one cluster near the \mf\ critical point\cite{quake} is
\begin{equation}
\label{densityfc}
\phi_{\rm cl} \sim {\epsilon^{1/2}\over R^d\epsilon^{2-d/2}} .
\end{equation}
This prediction has been
verified numerically\cite{RK1,frank}.

Are these clusters the critical phenomena fluctuations? To answer
this question we calculate the density of \mf\ fluctuations. The
partition function for the Gaussian approximation of the $\phi^{4}$
model is given by\cite{ma}
\begin{equation}
\label{phi4}
Z = \! \int \! \delta\phi\, \exp \bigl \{-\beta \! \int \! d{\vec
x}\, R^d[(\nabla\phi({\vec x}))^{2} + 
\epsilon\phi^{2}({\vec x})] \bigr\},
\end{equation}
where $\beta = 1/k_B T$ and $k_B$ is Boltzmann's constant. Because
we are interested only in the scaling properties of the
fluctuations, we take the order parameter
$\phi({\vec x})$ to be a constant $\phi$ over a volume
$\xi^d$ and zero outside, and obtain 
$Z \sim \! \int \! \delta\phi\, e^{-\beta
R^d\epsilon^{1-d/2}\phi^{2}}$. If we integrate $\phi$ until the
argument of the exponent becomes order unity, we find that the
density of the critical phenomena fluctuations scales as
\begin{equation}
\label{densitycf}
\phi\sim {\epsilon^{1/2}\over (R^d\epsilon^{2-d/2})^{1/2}} .
\end{equation}
That is, the structures with the density in
Eq.~(\ref{densitycf}) have a free energy cost of one. As in
the non-mean-field case, we expect these objects to be
the critical fluctuations.

{}From Eqs.~(\ref{densityfc}) and (\ref{densitycf}) we see that the
critical phenomena fluctuations are not the clusters,
but are considerably denser in the
\mf\ limit $R^d\epsilon^{2-d/2}>>1$. The
structure of the critical phenomena fluctuations in
\mf\ systems is that the ``vacuum'' is not featureless but contains
a very large number ($R^d\epsilon^{2-d/2}$ in a volume $\xi^d$) of
clusters. At the critical point (magnetic field $h=0$ in Ising
models), the mean number of up and down clusters is equal giving a
zero mean magnetization. Because these clusters are
independent\cite{KF,CK}, the fluctuations in the number of clusters
is order
$(R^d\epsilon^{2-d/2})^{1/2}$. These fluctuations make up the zero
magnetization background and are the critical phenomena
fluctuations. If we use Eq.~(\ref{densitycf}),
$\chi_T \sim \phi^2 \, \xi^d \sim \epsilon
R^d\epsilon^{-d/2}/R^d\epsilon^{2-d/2} =
\epsilon^{-1}$ as expected.

These arguments can be extended to spinodals in
\mf\ Ising models. If the spinodal is approached by varying $h$
and keeping $T$ fixed, the density of the clusters
is 
$\Delta h^{1/2}/R^d\Delta h^{3/2-d/4}$ with $\Delta h =
h_s-h$\cite{unger}. The numerator represents the magnetic field
scaling of the order parameter. The denominator is very large in
\mf\ and is the Ginsburg criterion for spinodals\cite{MK}. As at
the critical point, the mean number of up and down clusters is
equal when the infinite cluster, which is related to the metastable
magnetization, is subtracted. All scaling arguments for \mf\
critical points apply at spinodals with appropriate changes in the
values of the exponents ($\nu=1/4,
\alpha=1/2, \beta=1/2$, and $\gamma=1/2$ \cite{unger}).

We can determine the decay time of the clusters and the
critical phenomena fluctuations by constructing an action from the
linearized Langevin equation describing the dynamics of an Ising
model with long-range interactions\cite{ps}. If we assume a random
Gaussian noise, the probability measure for the order parameter
$\phi({\vec x},t)$ is\cite{kb}
\begin{eqnarray}
\label{sup}
\exp \biggl\{-\beta \! \int \! d{\vec x} \, dt \biggl [
\epsilon\bigl({\partial \tilde \phi({\vec x},t)\over \partial
t}\bigr)^{2} + (MR^d\epsilon^{2-d/2})^{2} \nonumber \\ 
\bigl\lbrace -\nabla^{2} \tilde \phi({\vec x},t) +
\epsilon \tilde \phi({\vec x}, t)\bigr\rbrace^{2} + 
H({\bar\psi},\psi)\biggr ] \biggr\},
\end{eqnarray}
where $M$ is a (constant) mobility, $\tilde \phi = \e^{1/2} \phi$,
and $H({\bar \psi},\psi)=
{\bar \psi}({\vec x},t)\bigl\lbrack {\partial\over \partial
t}+MR^d\epsilon^{2-d/2}\bigl( -\nabla^{2} +
\epsilon\bigr\rbrace\bigr)\psi({\vec x},t)$. The variables
${\bar\psi}({\vec x},t)$ and $\psi({\vec x},t)$ obey a Grassmann
algebra\cite{ps,kb} and can be used to convert the average over
the Gaussian noise to an average over the order
parameter\cite{ps}. In the \mf\ limit, the Langevin equation
describing the dynamics is linear, and the Grassmann fields and the
order parameter are independent\cite{kb}. Hence, the Grassmann
fields are irrelevant to a calculation of averages of functions of
$\phi$. Using the action in Eq.~(\ref{sup}), we can average a
function of $\phi$ by functionally integrating
$\phi$ up to its value at the critical point where we expect that
the argument of the exponential in Eq.~(\ref{sup}) is order one.
Because all of the terms in the action in Eq.~(\ref{sup}) are real
and positive, each term in the exponential must be order one,
which implies that $\epsilon \,\xi^d {\tilde \phi}^{2}/t \sim 1$.
Hence, the time scale that an object lives depends on
$\tilde \phi$.

For critical fluctuations, $\tilde \phi \sim
(R^d\epsilon^{2-d/2})^{-{1/2}}$, which leads to
$\tau
\sim
\epsilon^{-1}$, the scaling for the decorrelation time in a
\mf\ system with a non-conserved order parameter. For clusters,
$\tilde \phi_{\rm cl} \sim (R^d\epsilon^{2-d/2})^{-1}$, which leads
to 
$\tau_{\rm cl} \sim \epsilon^{-1}/R^d\epsilon^{2-d/2}$. We stress
that these arguments are valid only in the
\mf\ limit and are a good approximation in the \nmf\ case.

These considerations imply that the clusters have a
lifetime that is considerably shorter than the lifetime of the
critical phenomena fluctuations. In the \mf\ limit, 
$R^d\epsilon^{2-d/2}\rightarrow \infty$, the lifetime of the
clusters is zero.

We now apply these ideas to a \mf\ model of a supercooled fluid. We
consider the step potential $u(\gamma r)=0$ for
$\gamma r>1$ and
$u(\gamma r)=1$ for
$\gamma r\leq 1$\cite{ramos}. For $\gamma \to 0$,
$S(k)= 1/[1+\rho \beta u(k)]$ in the fluid phase \cite{gk}, where
$\rho$ is the density and $u(k)$ is the Fourier transform of
$\gamma^d u(x)$ with $x$ and
$k^{-1}$ scaled by $\gamma$. This \mf\ form of $S(k)$ indicates
that the system is unstable for $\rho \beta$ such that
$1+\rho \beta_s u(k_0)=0$, where $u(k_0)<0$ is the minimum of
$u(k)$. As $T \to
T_{s} = (k_B \beta_s)^{-1}$, $S(k_0)\sim
\epsilon^{-1}$\cite{gk}, where
$\epsilon=|T-T_{s}|/T_s$. The divergence of $S(k_0)$ is unchanged
if a short-ranged reference potential is added.

The scaling laws for the spinodal can be rewritten with $\epsilon$
as the scaling variable\cite{kletal} with $\nu=1/2$,
$2-\alpha=3$,
$\beta=1$, and $\gamma=1$ rather than the values quoted previously
for the $\Delta h$ scaling field\cite{proper}. The number of
clusters in a volume $\xi^d$ scales as
$R^d\epsilon^{3-d/2}$, the density of clusters as
$\epsilon/R^d\epsilon^{3-d/2}$, and the density of critical
phenomena fluctuations as 
$\epsilon/(R^d\epsilon^{3-d/2})^{1/2}$. The Ginsburg criterion is
$\Lambda = R^d \epsilon^{3-d/2} >> 1$, and the time scale for the
critical fluctuations and the clusters is
$\epsilon^{-1}$ and
$\epsilon^{-1}/R^d\epsilon^{3-d/2}$, respectively.

The primary difference between the Ising and fluid
spinodals is that $S(k)$ diverges at
$k_0 \neq 0$ for the fluid\cite{gk}. The Fourier transform of $S(k)$
yields a correlation function $\Gamma(r)$ that scales as
$r^{-1}e^{-r/\xi}\sin k_0 r$ in $d=3$, implying that the critical
fluctuations near the spinodal have a characteristic length
$k_0^{-1}<<\xi$ on which there is a periodic spatial variation.
Since the critical fluctuations are an incoherent superposition
of $\Lambda$ overlapping clusters, the spatial symmetry breaking
reflected in $\Gamma(r)$ and
$S(k)$ also occurs for the clusters. We can
show from an analysis of the Langevin equation that the
clusters have a triangular structure in $d=2$ and a bcc or layered
triangular structure in
$d=3$\cite{kletal}. (Nucleation near the spinodal is a
coalescence of clusters\cite{MK} and the
nucleation droplets have the symmetries discussed
above\cite{KL,yang}.)

To understand the consequences of our interpretation of \mf\ and
\nmf\ fluctuations, we obtain the $T$-dependence of $S(k_0)$
for a fluid using scaling arguments similar to the ones
used above.
The structure function is related to an integral over the
particles in a cluster times the probability that two
particles belong to the same cluster within a critical
fluctuation. The integral involves a phase factor $\sum_{j}
\exp[i{\vec k}_{0,j} \cdot{\vec r_j}]$, where
$|{\vec k}_{0,j}|=k_0$ and the index
$j$ labels the basis reciprocal lattice directions for the
indicated symmetry. This phase factor is multiplied by the
probability that a site belongs to a critical phenomena
fluctuation, $\epsilon/(R^d \epsilon^{3 - d/2})^{1/2}$, times the
probability that the second particle belongs to the same cluster,
$\e/(R^d \e^{3 - d/2})$, times the number of clusters, $(R^d \e^{3
- d/2})^{1/2}$. In addition, we average over the angles
corresponding to the random orientations of the clusters. These
considerations yield the scaling form:
\begin{eqnarray}
\label{sfscal}
S(k_0)&\sim& \! \int \! d{\vec k}^{\prime}_0 \, d{\vec
r}{\epsilon^{2}\over R^d\epsilon^{3-d/2}} e^{i \vec k_0 \cdot{\vec
r}} \nonumber \\
&\sim& {\epsilon^{2}\over
R^d\epsilon^{3-d/2}} \xi \sim R^{1-d} \epsilon^{-(3-d)/2} .
\end{eqnarray}
The spatial integral is over a region the size of
$\xi^d$ and $d{\vec k}^{\prime}_0$ denotes an integral
over angles.

Eq.~(\ref{sfscal}) predicts that
$S(k_0) \sim
\epsilon^{-\tilde \gamma}$ with $\tilde \gamma=1$ in
$d=1$, $\tilde \gamma=1/2$ in
$d=2$, and $\tilde \gamma=0$ in $d\geq 3$ in contrast to
$\gamma=1$ for all
$d$ as predicted by \mf\ theory\cite{gk}. We stress that the
weakening of the divergence of $S(k_0)$ in $d=2$ and its supression
for $d >2$ is a consequence of the quenched periodic structure of
the clusters. If this structure is modified by finite size effects
or defects, then the suppression might not be as strong.

The difference between the two calculations for the scaling
behavior of $S(k_0)$ is the limiting procedure. If
the limit
$\Lambda \to \infty$ ia taken {\it before} the calculation of
$S(k_0)$, the lifetime of the clusters is
zero in comparison to a measurement time. In this limit the
clusters appear rotationally symmetric and a
calculation similar to Eq.~(\ref{sfscal}) would yield $S(k_0) \sim
\epsilon^{-1}$ for all $d$, the same result as in Ref.~\cite{gk}.
However, if we assume 
$\l$ to be arbitrarily large but finite, the lifetime of the
clusters is nonzero and the measurement time is smaller than the
cluster lifetime. This assumption is consistent with the way
measurements are done in experiments and simulations.

It is difficult to estimate critical exponents by
fitting data directly to the suggested asymptotic form.
However, because the spinodal is well defined only in
the mean-field limit and simulations can be done only for finite
$R$, we must estimate
$\tg$ by approaching the pseudospinodal. But we cannot
approach it too closely because we will reach the
Becker-D\"oring limit where the system nucleates very quickly.
Because the Metropolis algorithm becomes very inefficient in
\nmf\ because almost all single particle moves are accepted, we
included a small hard core of diameter
$\s$ to decrease the acceptance probability.

Our Monte Carlo results (see Fig.~1) in $d=1$ were fit to
$AT^{\tg}\e^{-\tg}$ with $\tg = 1.0 \pm 0.05$ and $T_s \approx
0.83$. The $T^{\tg}$ term accounts in part
for corrections to scaling which typically depend on
$d$. The errors were estimated from different
possible fits. The $d=2$ results were fit to
$A\e^{-\tg}$ with $\tg = 0.4 \pm 0.05$ and $T_s \approx 0.81$. The
quality of the fits is not as good as in
$d=1$, but the estimate for $\tg$ is consistent with our
prediction of $\tg = 0.5$. The $d=3$ results are consistent
with $A\e^{-\tg}$ with $\tg = 0.16 \pm 0.02$ and
$T_s\approx 0.71$ and with $A \log\e + B$ with $T_s
\approx 0.86$. Details of the simulations will be discussed in a
longer paper\cite{kletal}.

\begin{figure}[tbp]
\hbox to\hsize{\epsfxsize=1.0\hsize\hfill\epsfbox{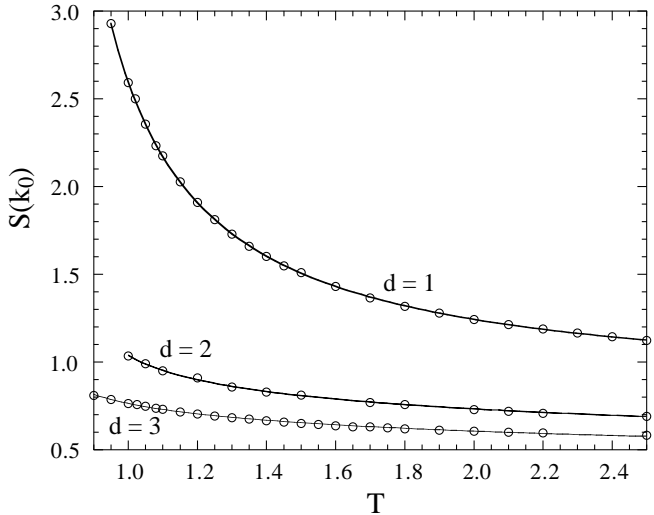}\hfill}
\caption{The $T$-dependence of $S(k_0)$ in $d=1$, 2, and 3 for the
step potential discussed in the text. The
$d=1$ results are for $N=10^4$ particles, interaction range $R=48$,
hard core diameter $\sigma=0.05$, and density $\rho = 1.95$. The
fit shown for $S(k_0)$ is to
$AT^{\tg}\e^{-\tg}$ with $\tg = 1.02$ and $T_s = 0.82$. For
comparison, $T_s = 0.85$ in the $R
\to \infty$ limit. The
$d=2$ results are for $N=10^3$, $R=6$, 
$\s = 0.15$, and $\rho = 1.95$. The fit for $S(k_0)$ is
$A\e^{-\tg}$ with $\tg = 0.44$ and $T_s = 0.81$. For comparison,
$T_s = 0.80$ for $R \to \infty$. The number of Monte
Carlo steps per particle (MCS) for each value of $T$ is $10^5$ in
$d=1$ and 2. The $d=3$ results are for $N=10^3$, $R=3$, $\sigma =
0.45$, and
$\rho = 1.95$ and fit equally well to $A\e^{-\tg}$ with
$\tg = 0.16$ and $T_s = 0.71$ and to $A\log \e + B$ with $T_s =
0.86$; $T_s = 0.70$ for $R \to \infty$. The two fits are
indistinguishable in the figure. The number of MCS for each data
point is between
$2
\times 10^5$ and $4 \times 10^5$.}
\end{figure}

In summary, our theoretical and simulation results imply that the
structure of the fluctuations in \mf\ and \nmf\ systems differs
qualitatively from that of non-\mf\ systems. This
structure leads to a suppression of the divergence of the measured
static structure function near a pseudospinodal relative to the
\mf\ prediction in systems for which a spatial symmetry is broken.
The dependence on
$d$ of this suppression is such that there is no apparent
divergence for
$d \geq 3$ subject to logarthmic corrections. In these
systems there is a growing correlation length as the
pseudospinodal is approached that cannot be obtained from a
measurement of $S(k)$. We note that a divergent
{\it dynamical} length has been found above the glass transition in
the spherical $p$ spin model
\cite{glotzer}. This work, coupled with our results, raises the
question of how the spinodal influences the dynamical length. Our
predictions have important implications for the understanding of
processes such as nucleation and glass formation in supercooled
fluids. In particular, we expect that the existence of clusters
should help us understand the universal scaling behavior found for
the dielectric response of organic glass formers\cite{dielectric}.

\medskip We thank R.\ C.\ Brower, Bulbul Chakraborty, and John
Rundle for useful discussions. This work was supported by NSF grant
DMR-9633385 (WK and HG) and DOE DE-FG02-95ER 14498 (WK and MA).
Acknowledgment is made to the donors of the Petroleum Research
Foundation, administrated by the American Chemical Society, for
partial support of the research at Kalamazoo College. Work on
LA-UR 00-117 at Los Alamos National Laboratory was supported by
the U.S.\ DOE LDRD-DR 98605.


\begin{references}
\bibitem{proper}The proper scaling field for the
spinodal is $a\Delta h + b\epsilon$, where $a$ and $b$ are
constants.
\end{references}
\end{document}